\documentclass[a4paper,fleqn,usenatbib]{mnras}

\usepackage[T1]{fontenc}
\usepackage{ae,aecompl}

\usepackage{graphicx}	
\usepackage{amsmath}	
\usepackage{amssymb}	

\title[A closer look at PSOJ0147]{A closer look at the quadruply lensed quasar PSOJ0147: spectroscopic redshifts and microlensing effect}

\author[C.-H. Lee]{
Chien-Hsiu Lee,$^{1}$\thanks{E-mail: leech@naoj.org}
\\
$^{1}$Subaru Telescope, National Astronomical Observatory of Japan, 650 North A'ohoku Place, Hilo, HI 96720, USA\\
}

\date{Accepted XXX. Received YYY; in original form ZZZ}

\pubyear{2017}

\begin{document}
\label{firstpage}
\pagerange{\pageref{firstpage}--\pageref{lastpage}}
\maketitle

\begin{abstract}
  I present timely spectroscopic follow-up of the newly discovered, quadruply lensed
  quasar PSOJ0147 from the Pan-STARRS 1 survey. The newly acquired optical spectra
  with GMOS onboard the Gemini north telescope allow us to pin down the redshifts of both the foreground
  lensing galaxy and the background lensed quasar to be z=0.572 and 2.341, providing firm basis for cosmography with future
  high cadence photometric monitoring.
  I also inspect difference spectra from two
  of the quasar images, revealing the microlensing effect.
  Long-term spectroscopic follow-ups will shed lights on the structure of the AGN and its environment. 
\end{abstract}

\begin{keywords}
gravitational lensing: strong -- quasars: general -- cosmology: observation
\end{keywords}


\section{Introduction} \label{sec:intro}
In the context of precision cosmology, determine local value of H$_0$ to percent level and compare it
  with CMB results provide tremendous leverage on the dark energy, neutrino physics,
  and the geometry of the Universe. Taking advantage of the distance ladder, the SH0ES
  team obtained a local measurement of H$_0$ to 2.4\% \citep{2016ApJ...826...56R} that is in
  tension (at 3.4$\sigma$ level)
with the Planck CMB results. Before claiming such tension is due to systematics in
the Planck measurement, or hints to new physics, it is important to have independent local H$_0$
measurements with comparable precision.

In this regard, the gravitational lensing time-delay method is an ideal alternative.
  This is because in strong lensing, lights from a multiply images source will go through different paths (geodesics). Since light travel times (or time-delays) differ proportionally to the space-time curvature, we can thus estimate the H$_0$. \cite{1964MNRAS.128..307R} originally suggested to use standard candles such as supernovae to measure the time-delay.
Supernovae, especially SNe Ia, have several advantages in time-delay measurement \citep[see e.g.][]{2017NatAs...1E.155L}.
First of all, SNe Ia have a distinct light curve shape, so we can measure the time-delay accurately if they
are discovered and followed-up well before the light maximum. Secondly, from the empirical light
curve template, we know the intrinsic brightness of the SNe, which provides the \textit{magnification
factor} that breaks the mass-sheet degeneracy and yields a direct measurement of H$_0$ \citep{2003MNRAS.338L..25O}.
However, due to the lack of all-sky transient surveys, the first multiply-lensed SN Ia, iPTF16geu, was only reported
very recently \citep{2017Sci...356..291G}. The discovery of iPTF16geu demonstrated that it is very difficult to establish the lensing
nature before the supernova light maximum, making it difficult to measure time-delay. To make it worse,
lensed SNe Ia will also suffer from microlensing effects \citep{2017ApJ...835L..25M}, further hampers accurate determination
of the time delay. In this regard, instead of SNe Ia, time-delay measurements were carried out with
quasars as background source. For example, the most recent time-delay measurements from the H0LiCOW team \citep{2017MNRAS.468.2590S} demonstrated
that it is possible to measure time-delays with 3 multiply-lensed quasars, enabling an H$_0$ estimate at 3.8\% level \citep{2017MNRAS.465.4914B}. This is possible because they also constrained the mass-sheet effect with spectroscopic observations \citep{2017MNRAS.467.4220R}, thus were able to break the constrain the effect of external masses along the line of sight. 

While the current time-delay method provides an consistent H$_0$ estimate as the distance ladder method, we still need to obtain H$_0$
to 1\% level, both to understand the tension with the CMB results from the Planck satellite, as well as to shed lights on the dark energy. However, at present there are only a handful of
suitable multiply-lensed systems, hence we need to increase the sample of suitable lensed quasars. Recently, \cite{2017ApJ...844...90B}
have identified a new quadruply lensed quasar candidates from the Pan-STARRS archive data. This system is of particular interests because the four
lensed quasar images are bright and provide a potentially suitable case for accurate time-delay measurement. This is especially the case when we consider to adopt the new high-cadence monitoring method proposed by \cite{2017arXiv170609424C}.

In this work I present timely follow-up spectra of this new candidate, with the aim to provide accurate redshifts of both the foreground lensing galaxy and the background quasar for cosmography. This paper is organized as follows:
the observation is documented in section \ref{sec:obs}. The analysis and results are presented in
section \ref{sec:res}.
I give the conclusion in section \ref{sec:con}.

\section{Observation} \label{sec:obs}

PSOJ0147 was first reported by \cite{2017ApJ...844...90B}, but was invisible for immediate follow-up at that time, and \cite{2017ApJ...844...90B} only derived the redshifts of the foreground lens and the background quasar photometrically.

\begin{figure}
  \centering
  \includegraphics[scale=0.4]{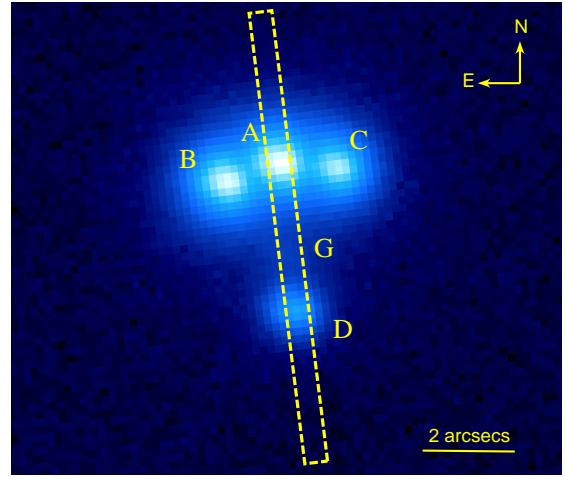}
  \caption{Illustration of the GMOS-N spectroscopic follow-up. The lensed quasar images are
    marked in A-D, starting with the brightest image A (i=15.40 mag). The 0.5arcsec-slit is shown
    in violet rectangle. With a single slit position (position angle = 7 degrees), we can obtain spectra
    of the brightest and faintest quasar image, as well as the lensing galaxy at once. The background image is a 10-sec acquisition exposure in the i-band from GMOS-N, showing a region of 10$\times$10 arcsec$^2$, where north is up and east is to the left.}
  \label{fig.slit}%
\end{figure}

While the lensed images are rather bright, with i=15.40-17.74 mag, the foreground lens is rather
faint, with i=19.5 mag.
Though \cite{2017arXiv170705873R} carried out spectroscopic observations of the lensed quasars with
the Keck Cosmic Web Imager (KWCI) onboard the Keck II telescope, they only obtained a total
integration time of $\sim$1200 seconds, not deep enough to detect the foreground lensing galaxy.
Further more, \cite{2017arXiv170705873R} only covered the blue side of the optical spectrum, hence their
redshift estimate of z=2.377 is heavily influenced by the BAL feature. Indeed, broader wavelength coverage
from the 2.5m Nordic Optical Telescope \citep{2017A&A...605L...8L} provides more accurate redshift estimate using
the cleaner forbidden line [C III], results in a smaller quasar redshift at z=2.341$\pm$0.001. However,
neither \cite{2017arXiv170705873R} nor \cite{2017A&A...605L...8L} were able to detect the foreground lens spectroscopically.
In this regard, I thus made use of the fast turnaround program of the
8-m Gemini North Telescope at Maunakea, with its Gemini Multi Object Spectrograph (GMOS-N).
To reveal the spectral feature of the faint foreground lensing galaxy, I obtain 4 $\times$ 1200-sec exposures with 0.5-arcsec slit to cover the foreground lensing galaxy, as well as the brightest and faintest quasar image (see Fig. \ref{fig.slit}).
The observations were carried out with the B600 grating to cover a wide wavelength
range in the optical. Given the photometric redshift of the foreground lens at z=0.57,
I expect to cover Ca H and K, as well as the G-band absorption features. In addition, with
the background quasar at z=2.341, I expect to cover the [C III] $\lambda\lambda$ 1908.734 Angstrom forbidden
line, to verify the redshift of the quasar as reported by \cite{2017A&A...605L...8L}.

The observations were carried out on September 2nd, 2017 with GMOS-N mounted on Gemini north
telescope (program ID: GN-2017B-FT-4). To avoid
bad pixels and remove cosmic rays, the observations were performed using an A-B-B-A sequence, with a 1 arcsec dithering. We took 1200-sec exposures at each dithering position, resulting in a total integration
time of 80 minutes.

Data reductions were carried out in a standard fashion using IRAF\footnote{IRAF is distributed by the National Optical Astronomy Observatory, which is operated by the Association of Universities for Research in Astronomy (AURA) under a cooperative agreement with the National Science Foundation.} and Gemini IRAF package\footnote{http://www.gemini.edu/node/11823}, including subtraction of bias,
flat fielding, calibrating the wavelength using a Thorium-Argon lamp, and calibrating the flux using a spectroscopic
standard star EG 131. 

\section{Results} \label{sec:res}

\begin{figure*}
  \centering
  \includegraphics[scale=1.4]{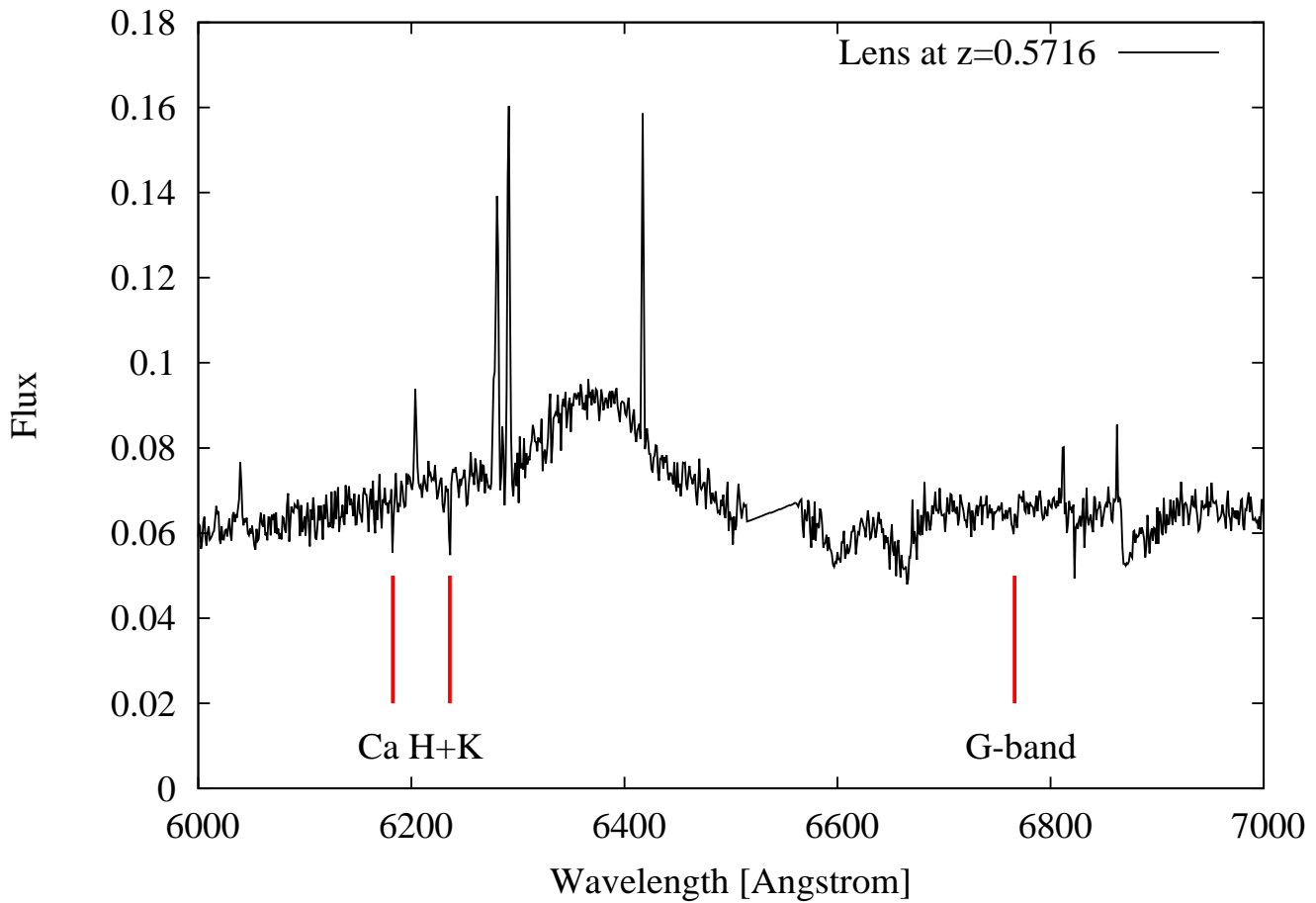}
  \caption{Spectrum of the foreground lensing galaxy from GMOS-N on-board the Gemini north telescope.
    Absorption features, such as Ca H$\lambda\lambda$ 3934, Ca K $\lambda\lambda$ 3969,
    and G-band $\lambda\lambda$ 4304 are marked with red labels,
    assuming a redshift of 0.5716. Please note that the lens galaxy and the lensed images of the background quasar are very close, thus the lens galaxy spectrum is blended by the light from the quasar images. Nevertheless this does not affect the redshift determination of the lens galaxy, as we can clear see the absorption lines from the lens galaxy}.
  \label{fig.spec}%
\end{figure*}

After data reduction, I can see traces of absorption lines from the foreground lensing galaxy,
e.g. Calcium H and K ($\lambda\lambda3934 \& 3969$) and G-band ($\lambda\lambda4304$).
The reduced lens spectrum, as well as these absorption features, as shown in Fig. \ref{fig.spec}.
From these absorption features, I am able to 
obtain the lens redshift; I use the \textit{splot} task in IRAF, and perform a Gaussian
fitting to the Ca H+K and G-band features. After measuring the Gaussian centroids from each images
I use their mean to determine the redshift, and
the standard deviation as the redshift uncertainty, resulting
in a redshift z=0.5716$\pm$0.0004. The values of the Gaussian centroid from
each of the absorption features are shown in Table \ref{tab.lens}.

\begin{table}
\caption{Gaussian centroid of absorption features in the lens spectrum.}
\centering
\begin{tabular}{lcc}
\hline\hline
& &  \\
Lines & $\lambda_{obs}$ [Angstrom] & z\\
\hline
Ca H $\lambda\lambda3934$ & 6182.68 & 0.5716\\
Ca K $\lambda\lambda3969$ & 6236.02 & 0.5712\\
G-band $\lambda\lambda4304$ & 6766.1 & 0.5720\\
\hline
\hline
\end{tabular}
\label{tab.lens}
\end{table}

\begin{figure*}
  \centering
  \includegraphics[scale=0.8]{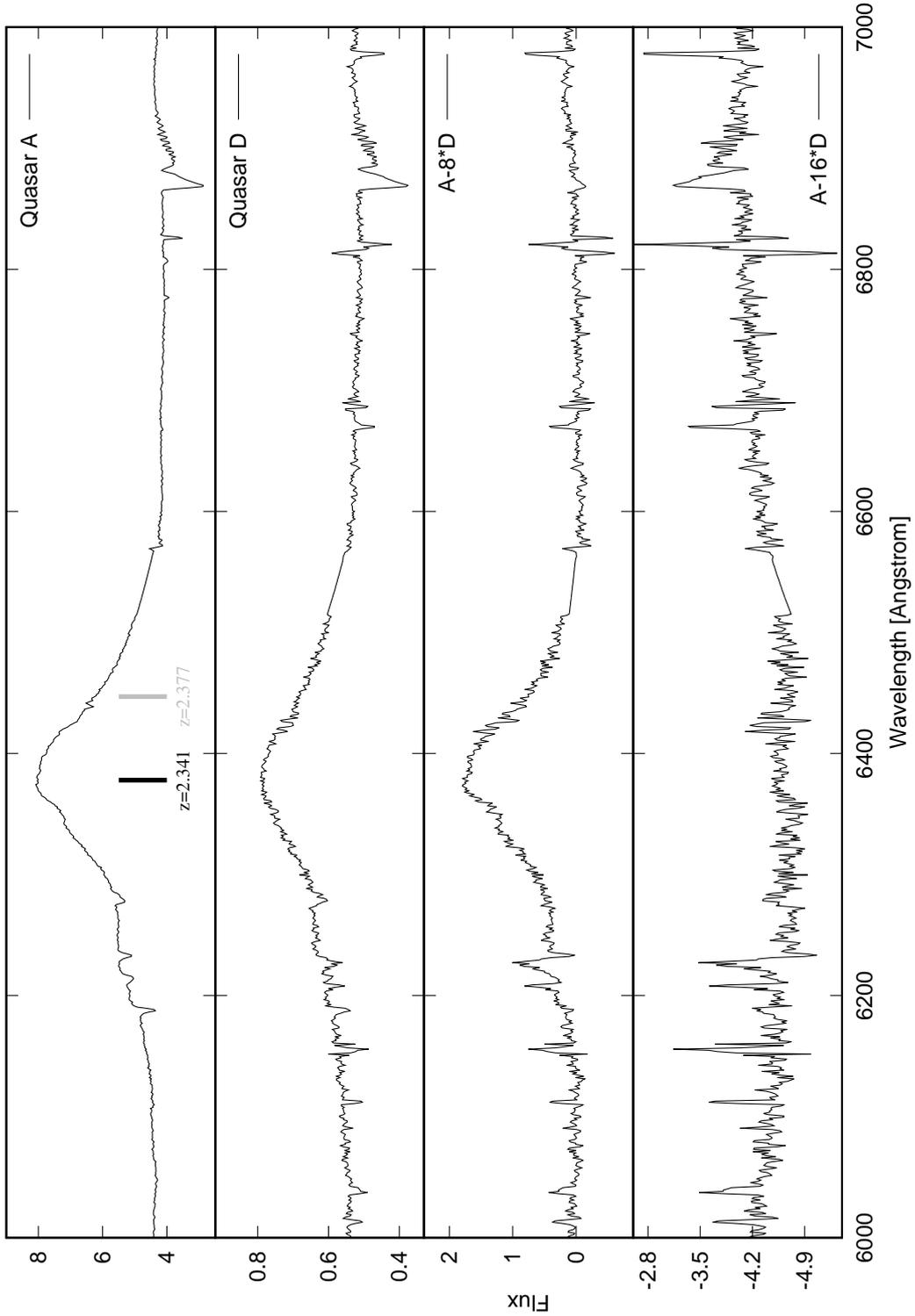}
  \caption{From top to bottom: 1) Spectrum of the brightest quasar image A, showing the [C III] emission line around 6380 Angstrom (marked by the black label), confirming the quasar to be z=2.341. If the quasar were at z=2.377 as Rubin et al. suggested, the [C III] emission line would be at 6447 Angstrom (marked by the gray lable), completely offset from the peak of the [C III] emission line shown in the plot; 2) Spectrum of the faintest quasar image D, with the [C III] emission line around 6380 Angstrom; 3) Difference spectrum of image A - 8$\times$ the flux of image D, nulling the flux contribution from the continuum and revealing only the flux contribution from [C III] emission line; 4) Difference spectrum of image A - 16$\times$ the flux of image D, nulling the flux contribution from the [C III] emission line and revealing only the flux contribution from continuum. I can see residual continuum bluewards of [C III] emission line, suggesting microlensing effect.}
  \label{fig.ratiospec}%
\end{figure*}

In addition to the lens, I also obtained spectra of the brightest and the faintest quasar images (A and D).
After reduction, I clearly see the [C III] forbidden emission line, as shown in Fig. \ref{fig.ratiospec}.
As [C III] is clean from BAL, it is ideal to be used to estimate redshifts. Indeed, \cite{2017A&A...605L...8L} reported
a quasar redshift of 2.341$\pm$0.001, contrasting to earlier results from \cite{2017arXiv170705873R}. As discussed
in \cite{2017A&A...605L...8L}, the differences may originate from the shorter wavelength coverage, as well as contaminations from
BALs in \cite{2017arXiv170705873R}. Here with the new Gemini/GMOS-N spectra, I again confirm [C III] of the background
quasar peaks at $\sim$ 6380 Angstrom, suggesting the background quasar is at z=2.341, in agreement with the results
from \cite{2017A&A...605L...8L} and smaller than the results from \cite{2017arXiv170705873R}.

With the spectra from quasar image A and D in hand, I can also investigate the microlensing effect spectroscopically
using difference spectrum. This has been done, for example, \cite{1993A&A...278L..15W} and \cite{2000A&A...364L..62L} have investigated
the difference spectra of the doubly lensed system HE1104-1805, and revealed the microlensing effects from spectra.
Furthermore, \cite{2012A&A...544A..62S} presented spectroscopic microlensing for 17 lensed quasars. If microlensing affects
the spectra, there will exist a factor $K$ where spectrum A - $K$ times spectrum D will cancel out the continuum and
I can only see the emission line. In addition, there will also exist another $K$ value where the emission lines will
be canceled out and leaves only blue residual continuum since microlensing affects most the central parts of the
accretion disk. By experimenting different values of $K$, with $K$ = 8 I am able to null the continuum and leave only the
emission features. With $K$ = 16, I can cancel out the emission feature, leaving a blue residual continuum.
The difference spectra are shown in Fig. \ref{fig.spec} as well.
These suggest we can see trace of the microlensing effects in PSOJ0147. The continuum is more susceptible to microlensing effect. Taking the flux ratio between the continuum of image A and D, I obtain a magnitude difference of 2.26 mag, close to the value of the observed quasar image magnitudes relative to image A, in particular r-band around 2.3 mag for image D vs. image A as shown in Fig. 7 of \cite{2017ApJ...844...90B}. On the other hand, the emission line flux ratio of image A and D suggests a magnitude difference of 3.01 mag, in the same direction (larger magnitude difference) as predicted by the lens modeling of \cite{2017ApJ...844...90B}.

\section{Conclusions} \label{sec:con}

PSOJ0147 is the first quadruply lensed quasar candidates identified in the Pan-STARRS archive data.
In this paper, I present follow-up spectra of PSOJ0147 using GMOS-N onboard the Gemini North Telescope.
The results can be summarized as follows:
\begin{enumerate}
  \item
    \textit{I provide the first spectrum of the foreground lensing galaxy.} 
    As the foreground lensing galaxy is very
    faint (i=19.5 mag), previous spectroscopic follow-up by \cite{2017arXiv170705873R} could not reveal the spectral signature of
    the foreground lens. With the power of the 8-m Gemini telescopes, and with a total integration time of 80 minutes, I
    was able to obtain spectrum of the foreground lensing galaxy for the first time. Using Ca H+K and G-band, I was able
    to determine the lens redshift to be z=0.5716, close to the photo-z prediction by \cite{2017ApJ...844...90B}. \\

\item
  \textit{I confirm the background quasar at a redshift of z=2.341}
  For cosmography studies, it is crucial to obtain accurate spectroscopic redshifts of both the foreground lens and the background quasar.
  Contrasting to the results of \cite{2017A&A...605L...8L}, previous study by \cite{2017arXiv170705873R} reported a larger redshift of the background
  quasar at z=2.377, which is likely to be affected by the BAL nature of the background quasar, and the narrow wavelength coverage
  (up to C IV at observers frame) in \cite{2017arXiv170705873R}. Here with the prominent [C III] emission line peaks at 6380 Angstrom,
  I confirm the background quasar at z=2.341. \\

\item
  \textit{I demonstrate the microlensing effect in difference spectra.} As microlensing affects most the central parts of the
  accretion disk, the continuum and the emission component of the background quasar will experience different lensing factor.
  I investigate the difference spectra between quasar image A and D, and show that with A - 8$\times$D we can null the continuum,
  while with A - 16$\times$D we can cancel out the emission line feature.
  Further epochs of spectroscopic monitoring will be
  crucial to study the environment around AGN of the background quasar.

\end{enumerate}

\section*{Acknowledgements}

I am indebted to the referee, whose comments greatly improved the manuscript.

Based on observations (GN-2017B-FT-4) obtained at the Gemini Observatory and processed using the Gemini IRAF package, which is operated by the Association of Universities for Research in Astronomy, Inc., under a cooperative agreement with the NSF on behalf of the Gemini partnership: the National Science Foundation (United States), the National Research Council (Canada), CONICYT (Chile), Ministerio de Ciencia, Tecnolog\'{i}a e Innovaci\'{o}n Productiva (Argentina), and Minist\'{e}rio da Ci\^{e}ncia, Tecnologia e Inova\c{c}\~{a}o (Brazil).

The author wish to recognize and acknowledge the very significant cultural role and reverence that the summit of Maunakea has always had within the indigenous Hawaiian community. I am most fortunate to have the opportunity to conduct observations from this mountain.








\bsp	
\label{lastpage}
\end{document}